\documentclass[12pt]{article}
\usepackage{lineno}
\usepackage[colorlinks, linkcolor=black, citecolor=black]{hyperref}
\usepackage{epsfig,epsf,psfrag,multirow,rotating}
\usepackage[latin1]{inputenc}
\usepackage{enumerate,enumitem,tikz}
\usepackage{lscape}
\usepackage{changepage,bbm,comment}
\usepackage{color,latexsym,graphicx}
\usepackage{amsfonts,amssymb,amscd,amsmath,amsbsy}
\usepackage{xcolor}
\usepackage{natbib}

\usepackage{mathtools}

\setlist[enumerate]{
  labelindent=0pt,
  leftmargin=*,
}

\setlist[itemize]{
  labelindent=0pt,
  leftmargin=*,
}

\usepackage{caption}
\usepackage{setspace} 

\usepackage{sectsty}
\sectionfont{\fontsize{12}{15}\selectfont}
\subsectionfont{\fontsize{12}{15}\selectfont}

\newtheorem{theorem}{Theorem}

\def\no{\noindent}

\newcommand{\Lfdr}{\textup{\mbox{Lfdr}}}

\renewcommand{\H}{\mbox{${\mathcal H}$}}
\definecolor{darkblue}{rgb}{0.0, 0.0, 0.5}

\newcommand*\circled[1]{\tikz[baseline=(char.base)]{\node[shape=circle,draw,inner sep=0.6pt] (char) {#1};}}

\mathtoolsset{showonlyrefs=true}

\begin{document}

{\centering {\large {\bf A powerful procedure that controls the false discovery rate with directional information}} \par}

\bigskip

\centerline{Zhaoyang Tian, \ Kun Liang \ and \ Pengfei Li\footnote{Zhaoyang Tian is doctoral student, Kun Liang is Associate Professor and Pengfei Li is Professor, Department of Statistics and Actuarial Science, University of Waterloo, Waterloo ON N2L 3G1,Canada (E-mails: {\em z26tian@uwaterloo.ca}, \ {\em kun.liang@uwaterloo.ca} \ and \ {\em pengfei.li@uwaterloo.ca}).}}

\bigskip

\bigskip

\hrule

{\small
\begin{quotation}
\no
In many multiple testing applications in genetics, the signs of test statistics provide useful directional information, such as whether genes are potentially up- or down-regulated between two experimental conditions. However, most existing procedures that control the false discovery rate (FDR) are $p$-value based and ignore such directional information. 
We introduce a novel procedure, the signed-knockoff procedure, to utilize the directional information and control the FDR in finite samples.
We demonstrate the power advantage of our procedure through simulation studies and two real applications. 

\vspace{0.3cm}

\no
\end{quotation}
}

\hrule

\bigskip

\bigskip

\section{Introduction}
In many modern statistical applications, hundreds or thousands of hypotheses are tested simultaneously.
For example, in microarray or RNA-seq studies aiming to find differentially expressed (DE) genes between diseased and healthy subjects, up to about twenty thousand hypotheses may be tested at the same time.
To adjust for multiplicity, we aim to control the false discovery rate (FDR), which is the expected proportion of false discoveries among rejections \citep{Ben95}.

Most existing procedures control the FDR based on $p$-values. 
Examples include the linear step-up procedure of \cite{Ben95}, the adaptive procedures of \cite{Storey04} and \citet{MacDonald17}, and many others.
In many practical applications, interesting findings have directions, whose information can be lost when using $p$-value based procedures.   
In the above example of finding DE genes between diseased and healthy subjects, the DE genes are either up- or down-regulated in the diseased subjects.   
In such a two-group comparison case, two-sample $t$-tests are commonly used, and the signs of the resulting $t$-statistics suggest whether the corresponding genes are potentially up- or down-regulated.  
Suppose that $t_i$ is the $t$-statistic for the $i$th gene under investigation, then the corresponding two-sided $p$-value is computed as $p_i=2\{1-F(|t_i|)\}$, where $F$ is the cumulative distribution function (cdf) of the $t$ distribution with an appropriate number of degrees of freedom.   
Only the absolute value of $t_i$ is used when computing $p_i$, and the directional information is lost in the $p$-value.

Thresholding on the $p$-values leads to symmetric rejection boundaries on the positive and negative sides of the original test statistics, such as the $t$-statistics in the example above.
By allowing asymmetric rejection boundaries, direction-adaptive procedures can improve power over traditional procedures.
Using $z$-values that preserve the directional information, \cite{Sun07} show that the optimal rule to maximize power with a constraint on the FDR is to set a cutoff on the local FDR, which is the posterior probability of a null hypothesis being true.
They show that under certain consistency conditions, their proposed data-adaptive procedure asymptotically controls the marginal FDR, which is a quantity closely related to the FDR.   
Their result requires, among other things, a consistent estimator of the true null proportion, which does not always exist in practice.
Suppose the target FDR level is $\alpha$, \cite{Orr14} suggest controlling the FDR at level $\alpha$ separately for $p$-values with positive and negative original test statistics and combining the rejections from both groups.
This is similar in spirit to the stratified FDR strategy proposed in \cite{Sun06}.
However, it is unclear whether the proposed procedure in \cite{Orr14} controls the FDR, and their simulation results show inflation of the realized FDR levels in certain settings.
Recently, \cite{Zhao16} propose a weighted $p$-value method by assigning different weights to $p$-values with different signs of the original test statistics.
The proposed weighting method is shown to asymptotically control the FDR when the true null proportion is estimated consistently or conservatively.
In summary, there are no existing direction-adaptive procedures that can control the FDR in finite samples.

Based on the idea of knockoffs from \citet{Barber15} and \citet{Lei18}, we introduce knockoffs to gain finite sample FDR control. 
First, we introduce the signed $p$-value to keep the directional information of the original test statistic. 
Then, knockoffs are constructed to mimic the behaviors of the signed $p$-values under null hypotheses.
Unordered pairs of the signed $p$-values and their knockoffs are given in the beginning instead of the original data, and the true identities of the knockoffs will be gradually revealed in later steps.
The number of knockoffs in the rejection region provides an empirical estimate of the number of false positives along the way, which allow us to control the FDR.

The rest of the paper is organized as follows.
In Section \ref{sec2}, we propose a novel procedure called the signed-knockoff procedure and show that it controls the FDR in finite samples.
In Section \ref{sec3}, through mixture modeling, we develop an algorithm to maximize power.
We numerically evaluate the proposed signed-knockoff procedure through simulation studies in Section \ref{sec4} and in real data applications in Section \ref{sec5}.  
Section \ref{sec6} concludes the paper.

\section{Signed-Knockoff Procedure}
\label{sec2}
First, we formally define the FDR.   Suppose there are $n$ null hypotheses being tested simultaneously: $H_{0i}$, $i=1,\,\cdots,\,n$.
Let $\H_0=\{i\text{: }H_{0i}\text{ is true}\}$ to denote the true null index set. The false discovery proportion (FDP) is defined as
\begin{equation*}
\text{FDP}=\frac{\#\{i:~i\in \H_0\ \&\ S_i=1\}}{\#\{i:~S_i=1\}\vee1},
\end{equation*}
where $S_i$ is the indicator variable of whether the $i$th null hypothesis is rejected and $a \vee b = \max(a,b)$.
The false discovery rate (FDR) is the expectation of the FDP, i.e., $\text{FDR}=\text{E}(\text{FDP})$.

For the ease of discussion, we will assume that the test statistics and thus the $p$-values are all continuously distributed.

\subsection{Signed p-values}
Suppose that $t_i$ is the test statistic for the $i$th null hypothesis, and $p_i$ is the corresponding $p$-value.   
To preserve the directional information in $t_i$, we could directly work on $t_i$ but will need to pay attention to its null distribution, which may vary from application to application.   
Alternatively, we can combine $\text{sign}(t_i)$ and $p_i$ to obtain a new statistic.
The new statistic has a one-to-one relationship with $t_i$ so information will not be lost in the transformation.\\

\noindent{\bf Definition (Signed $p$-value).} {\it For the $i$th null hypothesis with test statistic $t_i$ and $p$-value $p_i$, the signed $p$-value $q_i$ is defined as
\begin{equation*}
	q_i=\textup{sign}(t_i)(1-p_i).
\end{equation*}
}

We comment on some properties of the signed $p$-values.
First, the signed $p$-values range from $-1$ to $1$.
Next, under the null hypothesis, $\text{sign}(t_i)$ is independent of $p_i$.
Following from the fact that under the null hypothesis $p_i$ follows Uniform$(0,1)$, $q_i$ follows Uniform$(-1,1)$ if $i\in \H_0$.
Finally, as $p$-values under alternative hypotheses are more likely to be close to 0, the signed $p$-values under alternative hypotheses tend to be close to $-1$ and $1$.   
This is related to the reason we choose $(1-p_i)$ instead of $p_i$ in the definition of the signed $p$-value.   
If we have defined the signed $p$-value as $\text{sign}(t_i)p_i$, the signed $p$-value density under alternative hypotheses would not be continuous on the interval of $(-1,1)$.
Because of the distribution of the signed $p$-values under alternative hypotheses, we expect the rejection region of signed $p$-values to take the form of $[-1,a)\bigcup(b,1]$,
where $a$ is a negative number and $b$ is a positive number.

\subsection{The proposed procedure}
\label{skpsec}
Define the \emph{knockoff} of the $i$th signed $p$-value $q_i$ as 
\begin{equation*}
\tilde{q}_i=\text{sign}(q_i)-q_i.
\end{equation*}
As $p$-values are continuously distributed, no signed $p$-values should be exactly equal to $0$, so we do not consider how to set the values of knockoffs for zeros.
From the definition, $q_i$ and $\tilde{q}_i$ share the same signs.   If $q_i$ is positive, then $q_i$ and $\tilde{q}_i$ are symmetric about 1/2.   Similarly, if $q_i$ is negative, then $q_i$ and $\tilde{q}_i$ are symmetric about $-1/2$.   Furthermore, if the $i$th null hypothesis is true, $q_i$ and $\tilde{q}_i$ would both follow Uniform(-1,1).   In summary, under the null hypothesis, $\tilde{q}_i$ has the same distribution as $q_i$ and hence is a knockoff.

Before getting into the details of our proposed procedure, we describe the general idea behind it.   
Our procedure is a stepwise procedure.
We will start with an initial guess of the rejection region, which should be reasonably large.
In general, the less the number of rejections, the less the false discovery proportion.
Thus, we make the rejection region $R$ shrink towards the two endpoints $-1$ and $1$ gradually at each step by accepting one null hypothesis at a time. 
At each step, we estimate the FDR as
\begin{equation*}
\widehat{\text{FDR}}=\frac{1+\#\{i:~\tilde q_i\in R\}}{\#\{i:~q_i\in R\}\vee1}.
\end{equation*}
Because $\tilde q_i$ follows the same distribution as $q_i$ under the null hypothesis, 
the number of false discoveries $\#\{i:~i\in \H_0\ \&\ q_i\in R\}$ follows the same distribution as $\#\{i:~i\in \H_0\ \&\ \tilde q_i\in R\}$.
Thus, the numerator of $\widehat{\text{FDR}}$, $1+\#\{i:~\tilde q_i\in R\}$, provides a slight overestimate of the number of false discoveries.
$\widehat{\text{FDR}}$ tends to decrease after repeated shrinkages.   
We will stop the shrinkage at the first time $\widehat{\text{FDR}} \leq \alpha$ or when all null hypotheses have been accepted, whichever comes the first.

We will first define some notation.
We divide the signed $p$-values $\{q_i\}_{i=1}^n$ into two groups, $\{q_i^+\}_{i=1}^{n_+}$ and $\{q_i^-\}_{i=1}^{n_-}$, according to their signs.
Denote $q_{(i)}^+$ as the positive signed $p$-value which is the $i$th closest to $1/2$, and $q_{(j)}^-$ as the negative signed $p$-value which is the $j$th closest to $-1/2$.
Let $\alpha$ be the nominal FDR level.

For any rejection region and the $i$th null hypothesis, $i=1,\ldots,n$, we reject the null hypothesis if any element of its corresponding unordered pair $\{q_i,\tilde q_i\}$ falls into the rejection region.
After $k$ shrinkage steps in our procedure, let $i_k$ and $j_k$ be the number of accepted hypotheses on the positive side and negative side, respectively.
Then we can define the rejection region after $k$ shrinkage steps as $R_{k}=[-1,q^-_{(j_{k})}\wedge\tilde q^-_{(j_{k})})\bigcup(q^+_{(i_{k})}\vee\tilde q^+_{(i_{k})},1]$, where $a \wedge b = \min(a,b)$.

Let $\mathcal{I}_k=\{i:~q_i\in\{q^+_{(1)},...,q^+_{(i_k)},q^-_{(1)},...,q^-_{(j_k)}\}\}$ be the index set of accepted hypotheses after $k$ shrinkage steps, and $\mathcal{J}_k=\{i:~i=1,...,n\}/\mathcal{I}_k$ be the index set of rejected hypotheses after $k$ steps.
Let $b_i=I(|q_i|>1/2)$.
Define the $\sigma$-algebra $\mathcal{F}_k$ as
\begin{equation*}
\mathcal{F}_k=\sigma\Big(\{\text{min}(q_i,\tilde{q}_i)\}_{i=1}^n,\,\{b_i\}_{i\in\mathcal{I}_k}\Big).
\end{equation*}
Later in Step 2 of our procedure, we will require a decision to be $\mathcal{F}_{k}$-measurable.
In $\mathcal{F}_{k}$, $\{\text{min}(q_i,\tilde{q}_i)\}_{i=1}^n$ contains all the location information of $\{q_i,\tilde{q}_i\}$ pairs, and $b_i$ is the indicator of which element of the unordered pair $\{q_i,\tilde{q}_i\}$ is $q_i$.
Thus, $\mathcal{F}_{k}$ encapsulates the knowledge of all unordered pairs $\{q_i,\tilde{q}_i\}$'s of currently rejected null hypotheses plus the $q_i$ values of already accepted null hypotheses after the $k$th step.
In other words, for the currently unaccepted null hypotheses, the true identities of the signed $p$-values in the unordered pairs are \textit{masked} from us. Once we decide to accept a null hypothesis, the true identity of the corresponding signed $p$-value will be revealed to us.
Our proposed procedure is as follows.\\

\noindent{\bf Definition (Signed-knockoff procedure).} 
\begin{enumerate}[label ={\bf Step \arabic*}:]
\item Initialization.

Set the shrinkage step number $k=0$.
For convenience, we set $i_0=j_0=1$ and let $R_0$ be the initial rejection region.

\item Choice.

At the $(k+1)$th shrinkage step, we make the rejection region shrink to $R_{k+1}$ by moving one and only one signed $p$-value pair out of the rejection region.
More specifically, we choose one of the following:
\begin{enumerate}[label=\protect\circled{\arabic*}]
\item Proceed on the positive side: set $i_{k+1}=i_k+1$ and $j_{k+1}=j_k$.
\item Proceed on the negative side: set $i_{k+1}=i_k$ and $j_{k+1}=j_k+1$.
\end{enumerate}
As boundary conditions, we choose \circled{1} if $j_k=n_-$, and we choose \circled{2} if $i_k=n_+$.

If neither $i_k$ nor $j_k$ reaches its bound, we require the choice between \circled{1} and \circled{2} to be $\mathcal{F}_{k}$-measurable.

Increase $k$ by 1.

\item Stopping condition.

If
\begin{equation*}
\widehat{\text{FDR}}_k=\frac{1+\#\{i:~\tilde q_i\in R_k\}}{\#\{i:~q_i\in R_k\}\vee1}\leq\alpha,
\end{equation*}
or
\begin{equation*}
i_k=n_+\mbox{ and }j_k=n_-
\end{equation*}
we stop and reject all null hypotheses whose corresponding signed $p$-values are in $R_k$.
Otherwise, we return to Step 2.

\end{enumerate}

\subsection{FDR control}
We can see that the procedure above leaves out a key point, that is, how to decide whether to choose \circled{1} or \circled{2} in Step 2.
In fact, it can be arbitrarily chosen as long as the choices are $\mathcal{F}_k$-measurable, that is, the choices are made based on the partially masked information in $\mathcal{F}_k$.
We can show that regardless of the choices, the signed-knockoff procedure controls the FDR at the nominal level $\alpha$. 
We will assume the \emph{null independence condition} where the true null test statistics are independent of each other and independent of alternative test statistics.
A similar null independence condition imposed on $p$-values is commonly used in the literature, for example, see \cite{Ben95}, \cite{Storey04}, \cite{Liang12.FDR.Est}, and \citet{Lei18}, among many others.\\
\begin{theorem}
\label{thmfdr}
Under the null independence condition, if $p$-values under null hypotheses follow $\text{Uniform}(0,1)$ and are independent of the signs of test statistics, the signed-knockoff procedure controls the FDR at level $\alpha$.
\end{theorem}
In multiple testing applications where test statistics have signs, the true null test statistic distribution is typically symmetric about zero. Examples include the central $t$ distribution and the standard normal distribution. For such symmetric distributions, a true null test statistic has an equal probability to be positive or negative, and its sign is independent of the corresponding $p$-values. The proof of Theorem 1 is included in the supplementary material.

\section{Powerful choice}
\label{sec3}
\noindent
In Step 2 of our proposed procedure, the choices of which side to proceed on will not lead to the loss of FDR control, but they will affect power.
In the ideal case where the data generating model is completely known, it is well established in the Bayesian decision-theoretic framework that the local FDR is the optimal ranking statistic to maximize power with a constraint on the FDR, see \cite{Muller04}, \cite{Sun07}, and \citet{Lei18}.
This result implies that we should choose the pair whose corresponding null hypothesis is more likely to be true. 

If all signed $p$-values are available to us, we can consider the two-group model \citep{Efron01}:
\begin{equation*}
f(q) = \pi_0 f_0(q) + (1-\pi_0) f_1(q),
\end{equation*}
where $\pi_0$ is the probability of that the null hypothesis is true, $f_0$, $f_1$, and $f$ are the null, alternative, and overall densities of the signed $p$-values, respectively.   Under the null hypothesis, the signed $p$-value follows Uniform$(-1, 1)$, i.e., $f_0(q)=1/2$.   For the alternative density, we model it parametrically as a two-component mixture of transformed beta distributions.   More specifically, let
\begin{equation*}
f_1(q)=\lambda\frac{\alpha}{2}\left(\frac{q+1}{2}\right)^{\alpha-1}+(1-\lambda)\frac{\beta}{2}\left(\frac{1-q}{2}\right)^{\beta-1},
\end{equation*}
where $\lambda$ is the mixing proportion, and $\alpha$ and $\beta$ are the shape parameters of the component beta distributions.   The two components are regular beta distributions transformed from the range of $(0,1)$ to the range of $(-1,1)$.   We set one of the shape parameter in each beta distribution to 1 such that the alternative negative or positive signed $p$-value density is strictly decreasing or increasing, respectively, over the range of $(-1,1)$.

Recall that after $k$ shrinkage steps, in Step 2 of the signed-knockoff procedure, the decision needs to be made based on $\mathcal{F}_{k}$.
If the $i$th signed $p$-value has not been accepted, we only observe the masked unordered pair $\{q_i,\tilde{q}_i\}$.
The probability density function of $\{q_i,\tilde{q}_i\}$ should be the sum of those of $q_i$ and $\tilde{q_i}$, as the true signed $p$-value in $\{q_i,\tilde{q}_i\}$ can either be $q_i$ or $\tilde{q}_i$.
That is
\begin{align*}
f(\{q_i,\tilde{q}_i\}) = \pi_0+(1-\pi_0)f_1(q_i)+(1-\pi_0)f_1(\tilde{q}_i).
\end{align*}

Then the log-likelihood function after $k$ shrinkage steps is
\begin{align*}
l^{(k)}(\pi_0,\lambda,\alpha,\beta)=&\sum_{i\in\mathcal{I}_k}\log f(q_i) + \sum_{i\in\mathcal{J}_k}\log f(\{q_i,\tilde{q}_i\})\\
=&\sum_{i\in\mathcal{I}_k}\log\left\{\frac{\pi_0}{2}+(1-\pi_0)f_1(q_i;\lambda,\alpha,\beta)\right\}\\
&+\sum_{i\in\mathcal{J}_k}\log\Big\{\pi_0+(1-\pi_0)f_1(q_i;\lambda,\alpha,\beta)\\
&+(1-\pi_0)f_1(\tilde{q}_i;\lambda,\alpha,\beta)\Big\}.
\end{align*}
Maximizing the likelihood directly with respect to $\pi_0$, $\lambda$, $\alpha$, and $\beta$ can be numerically challenging.   Instead, we develop an EM algorithm by treating the true null statuses of all hypotheses and the true identities of $q_i$'s within the masked pairs as missing information.   The details of the EM algorithm are described in the supplementary material.   Through the EM algorithm we obtain the maximum likelihood estimators $\hat{\pi}_0$, $\hat{\lambda}$, $\hat{\alpha}$, and $\hat{\beta}$.      Finally, we compute the local FDR estimate for an unordered pair as
\begin{align*}
\Lfdr(\{q,\tilde{q}\}) = \frac{\hat{\pi}_0}{f(\{q,\tilde{q}\}; \hat{\pi}_0, \hat{\lambda}, \hat{\alpha}, \hat{\beta})}.
\end{align*}

After $k$ shrinkage steps and in Step 2 of the signed-knockoff procedure, we choose \circled{1} to proceed on the positive side if 
$$
\Lfdr(\{q^+_{(i_k+1)},\tilde{q}^+_{(i_k+1)}\}) \geq \Lfdr(\{q^-_{(j_k+1)},\tilde{q}^-_{(j_k+1)}\}),
$$
i.e., if the next positive pair is more likely to come from the true null hypothesis than the next negative pair. We will choose \circled{2} otherwise.

\section{Simulation studies}
\label{sec4}
In this section, we conduct simulation studies to compare the numerical performance of the proposed signed-knockoff procedure with other direction-adaptive procedures.   The power advantage of direction-adaptive procedures over $p$-value based methods has been demonstrated in the literature, see simulation results of \citet{Sun07} and \citet{Zhao16}.
Further evidence of this power advantage can also be found in the supplementary material.

\subsection{Candidate procedures}
\label{sec41}
We consider the following candidate procedures:
\begin{itemize}[label={--}]
\item \texttt{SK}, the signed-knockoff procedure.
\item \texttt{ORC}, the oracle procedure based on $z$-values proposed by \citet{Sun07}.   The local FDR for each hypothesis is computed using the true parameters. The FDR control is achieved by thresholding the local FDR such that the average local FDR below the threshold is close to but less than the target FDR level. 
\item \texttt{WBH}, the weighted BH procedure proposed by \citet{Zhao16}. As in their simulation study, we used the conservative least-slope $\pi_0$-estimator from \citet{Ben00}.
\end{itemize}

The implementation of \texttt{ORC} requires the knowledge of true parameters and is not practical in real applications.   However, \texttt{ORC} can provide a reference of the power upper bound in our simulation.
The data-adaptive procedure of \citet{Sun07} requires a consistent $\pi_0$-estimator, which does not exist in the setting of Section \ref{sec43}. 
For the same reason, we do not consider the performance of \texttt{WBH} with a consistent $\pi_0$-estimator.

The simulation results are based on 200 repeated runs, and we report the average FDR and power.

\subsection{Independent normal statistics}
\label{sec42}
The simulation setting here is similar to that in \citet{Sun07} and \citet{Zhao16}.   More specifically, we generate test statistics $z_i$ independently and identically from the following normal mixture model:
$$z_i\sim(1-p_1-p_2)\text{N}(0,1)+p_1\text{N}(\mu_1,1)+p_2\text{N}(\mu_2,1),$$
where $\mu_1<0<\mu_2$.
The test statistics from the null distribution follow the standard normal distribution.
Alternative test statistics follow the normal distribution with mean $\mu_1$ or $\mu_2$ and standard deviation $1$.
We emphasize that the generative model above differs from the beta model we used to compute local FDR estimates in Section \ref{sec3}. That is, our estimation model is misspecified.

We study the following three cases of parameter combinations:
\begin{enumerate}[label=(\alph*)]
\item Let $p_1$ vary between 0 and 0.2 and set $p_2=0.2-p_1$, $\mu_1=-3$ and $\mu_2=3$.
\item Let $p_1$ vary between 0 and 0.2 and set $p_2=0.2-p_1$, $\mu_1=-3$ and $\mu_2=6$.
\item Let $\mu_2$ vary between $2$ and $6$ and set $p_1=0.18$, $p_2=0.02$ and $\mu_1=-3$.
\end{enumerate}
Under each case, we test $n=5000$ hypotheses simultaneously and set the nominal FDR level $\alpha=0.1$.

\begin{figure}
\centerline{
\includegraphics[scale=0.9]{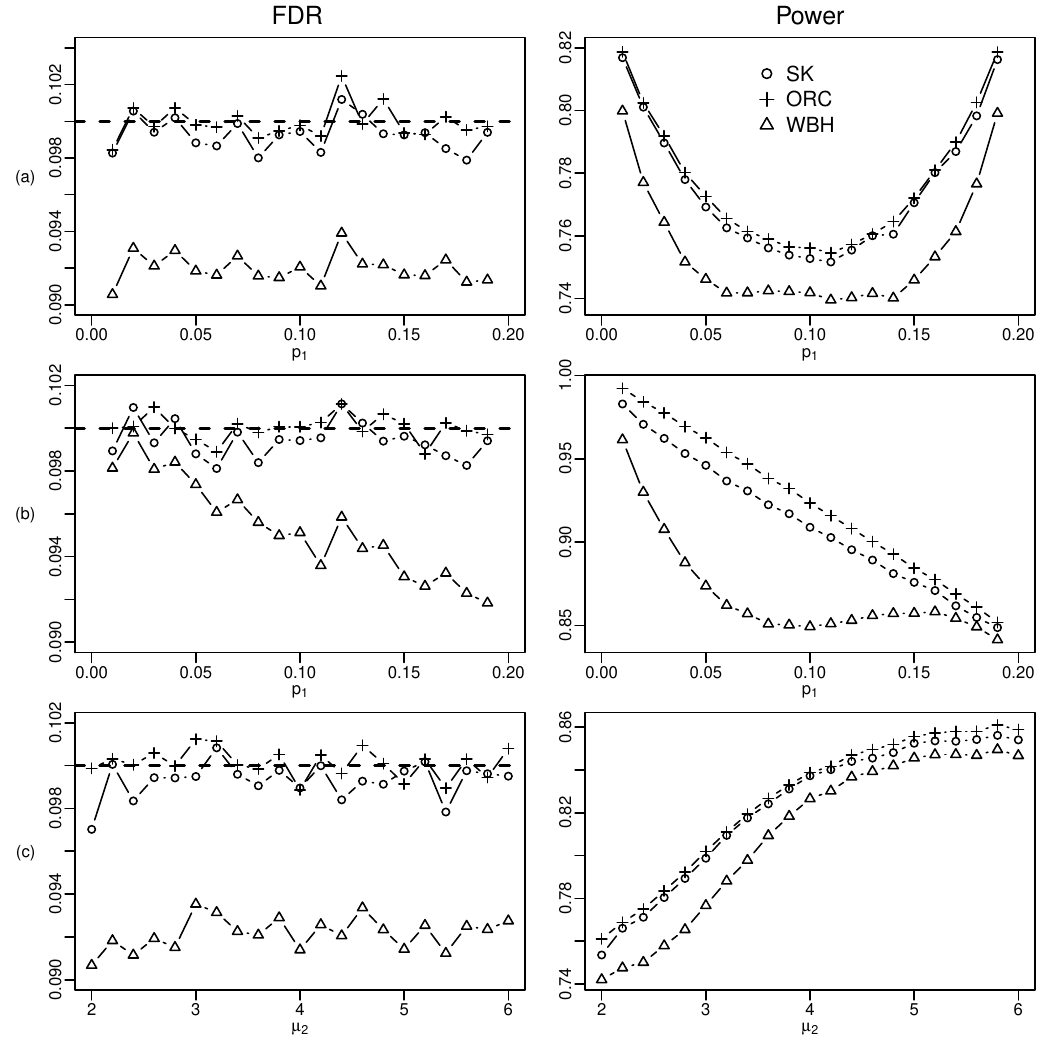}}
\caption{Realized FDR levels and power in the setting of independent normal statistics. The left column shows the realized FDR levels, and the right column shows the realized power. Rows (a)--(c) show the results in cases (a)--(c), respectively.}
\label{fdrfig1}
\end{figure}

The left column in Figure~\ref{fdrfig1} plots the average realized FDR levels.   
Across all three cases, we can see that the FDR levels of \texttt{SK} and \texttt{ORC} are quite close to each other and to the target level, and \texttt{SK} is slightly more conservative.   
\texttt{WBH} is the most conservative procedure among all candidates, and its FDR levels can be markedly lower than the target level.

The right column in Figure~\ref{fdrfig1} plots the average power levels.  
It is not a surprise that \texttt{ORC}, as the oracle method, is overall the most powerful procedure. 
However, its power advantage over \texttt{SK} is minor, especially in cases (a) and (c).   
Across all three cases, the power curves of \texttt{SK} mimic that of \texttt{ORC}.   
Like \texttt{ORC}, \texttt{SK} ranks the hypotheses by their local FDR.
Although we estimate the local FDR based on partially masked data and a misspecified parametric model, the realized power curves show a remarkably similar trend as \texttt{ORC}, whose local FDR estimates are computed from the true distribution.
Furthermore, \texttt{SK} is substantially more powerful than \texttt{WBH} in all cases.   
As shown in the left column of Figure~\ref{fdrfig1}, \texttt{WBH} is consistently more conservative than \texttt{SK}, but the power advantage of \texttt{SK} over \texttt{WBH} cannot be explained completely by \texttt{WBH}'s conservativeness.   
For example in case (b), the FDR curve and power curve of \texttt{WBH} relative to \texttt{SK} show quite different trends.
When $p_1 \leq 0.05$ in case (b), \texttt{WBH} and \texttt{SK} have comparable FDR levels, but \texttt{SK} is significantly more powerful than \texttt{WBH}.
Furthermore, even if a consistent $\pi_0$-estimator is used with \texttt{WBH}, \texttt{SK} still holds its power advantage over \texttt{WBH}, see simulation result in the supplementary material.

\subsection{Dependent $t$-statistics}
\label{sec43}
We evaluate the performance of candidate procedures with a dependent structure that mimics the microarray gene expression data.
The simulation setup is similar to the dependence simulation setting in \cite{Liang16}, and the sample size here is the same as the sample size in the first real data application in Section \ref{sec5}.
More specifically, we randomly divide the null hypotheses into blocks of size $b=20$.
Within each block, the gene expression levels of 6 subjects are generated independently according to $MVN(\underline{0}_b, \Sigma_{b \times b})$, where $\Sigma_{b\times b}=(\sigma_{ij})_{b\times b}=(\rho^{|i-j|})_{b\times b}$ with $\rho=-.7$.
That is, we consider an autoregressive order one (AR1) dependence within each block to mimic the mixing of positive and negative correlations among genes in the same biological pathway.   
Among the 6 subjects, 3 subjects are designated as the Treatment group, and the other 3 subjects form the Control group.
The DE genes have a probability of $p_1$ to be down-regulated in the Treatment group subjects, and we add a treatment effect $\mu_1$ to their gene expression levels.   
Similarly, the DE genes have a probability of $p_2$ to be up-regulated in the Treatment group subjects, and we add a treatment effect $\mu_2$ to their gene expression levels.
The test statistics $t_1, \ldots, t_n$ were computed by the regular two-sample $t$-test. 
Then we study three cases of parameter variations as in Section \ref{sec42}.

Unlike in Section \ref{sec42} where the test statistics follow normal distributions, here we have $t$-statistics, which are commonly used in gene expression studies and many other types of studies. 
As suggested by \citet{Sun07}, we could transform $t$-statistics to $z$-values such that the $z$-values from the true null hypotheses follow the standard normal distribution. More specifically, in our setting, we can compute $z_i=\Phi^{-1}(F(t_i))$, $i=1, \ldots, n$, where $\Phi$ is the cdf of the standard normal distribution and $F$ is the cdf of the central $t$ distribution with 4 degrees of freedom. However, the $z$-values from the alternative hypotheses do \textit{not} follow normal distributions. 
The data-adaptive procedure of \citet{Sun07} utilizes the $\pi_0$-estimator of \cite{Jin07}, which is consistent when all $z$-values follow normal distributions.
However, the $\pi_0$-estimator of \cite{Jin07} may not be consistent when the original test statistics are not normally distributed, such as the $t$-statistics in this setting. In the supplementary material, we show that \texttt{WBH} used with the $\pi_0$-estimator of \cite{Jin07} in this setting leads to inflated FDR levels.

\begin{figure}
\centerline{
\includegraphics[scale=0.9]{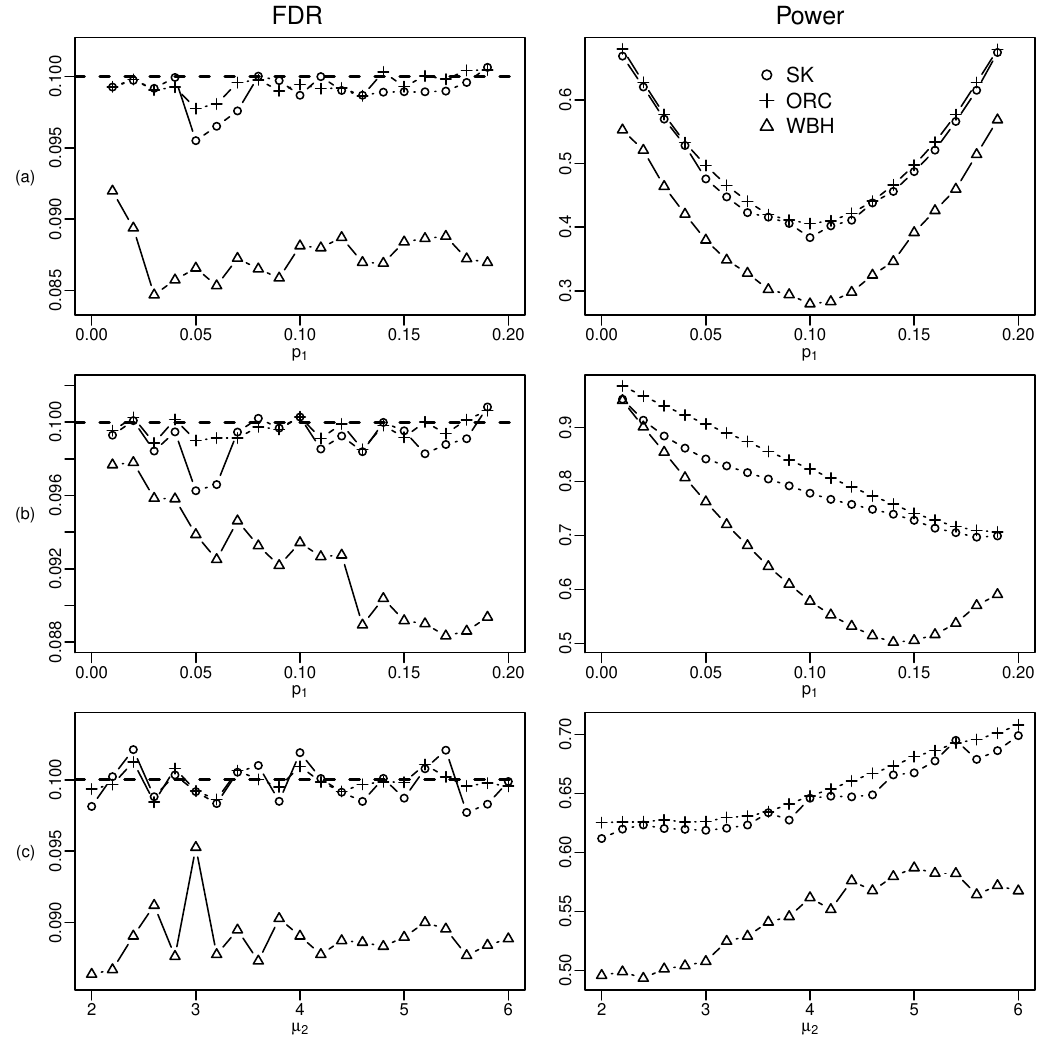}}
\caption{Realized FDR levels and power in the setting of dependent $t$-statistics. The left column shows the realized FDR levels, and the right column shows the realized power. Rows (a)--(c) show the results in cases (a)--(c), respectively.}
\label{fdrfig4}
\end{figure}

The average realized FDR and power levels are shown in the left column and the right column in Figure~\ref{fdrfig4}, respectively.
Across all three cases, there is no evidence suggesting that \texttt{SK}, \texttt{ORC} and \texttt{WBH} exceed the target FDR level.   
\texttt{ORC} is only slightly more powerful than \texttt{SK}, which is significantly more powerful than \texttt{WBH}.

We briefly summarize the simulation results under both independence and dependence settings. The performance of \texttt{SK} is close to the oracle, and \texttt{SK} offers significant power improvement over \texttt{WBH}.

\section{Real data applications}
\label{sec5}
We now illustrate the power advantage of our method through two real data applications.
The test statistic distribution in the first application is strongly asymmetric while the distribution in the second application is roughly symmetric about 0.

Except \texttt{ORC}, the oracle procedure proposed by \cite{Sun07}, all procedures compared in Section \ref{sec4} are considered: \texttt{SK}, our proposed signed-knockoff procedure; \texttt{WBH}, the weighted BH procedure by \cite{Zhao16}.   
For illustration purpose, we also considered a $p$-value based procedure that does not utilize the directional information: \texttt{RB}, the right-boundary procedure proposed by \cite{Liang12.FDR.Est}.
\citet{MacDonald17} show that \texttt{RB} controls the FDR in finite samples, and simulation results suggest \texttt{RB} is one of the most powerful $p$-value based adaptive procedures.  
We let the nominal FDR level $\alpha$ vary from 0.01 to 0.2, and focus on the result when $\alpha=0.1$.

\subsection{Thale cress seedlings}
\label{fdrsec5subsec1}
\cite{Jang14} conducted a microarray study to identify DE genes between two genotypes, the wild-type and the mutant type.
The expression levels of $n=22810$ genes are compared between the two genotypes, using 3 samples of tissues from each genotype.
We compute the regular two-sample $t$-statistics for all genes as $t_1, \ldots, t_n$.
The corresponding $p$-values are computed as $p_i=2\left\{1-F(|t_i|)\right\}$, $i=1, \ldots, n$, where $F$ is the cdf of the $t$ distribution with 4 degrees of freedom.

\begin{figure}
\centerline{
\includegraphics[width=6in]{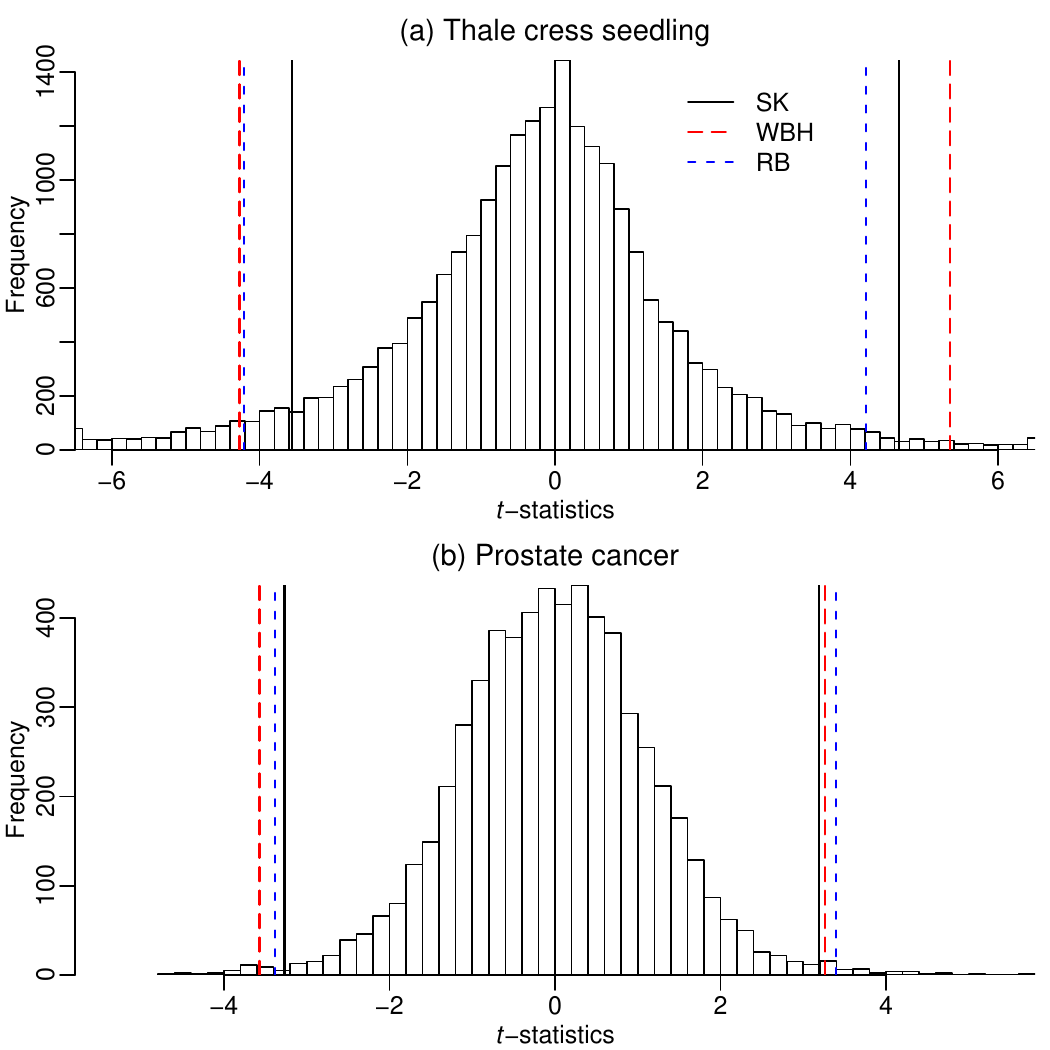}}
\caption{Histogram of $t$-statistics and boundaries of rejection regions given by different procedures when the nominal FDR level $\alpha=0.1$. Panel (a) shows test statistics from the thale cress seedling data analysis, and Panel (b) shows test statistics from the prostate cancer data analysis. As there are too many extreme $t$-statistics in the thale cress dataset, the histogram in Panel (a) is truncated.}
\label{fdrfig9}
\end{figure}

\begin{figure}
\centerline{
\includegraphics[width=6in]{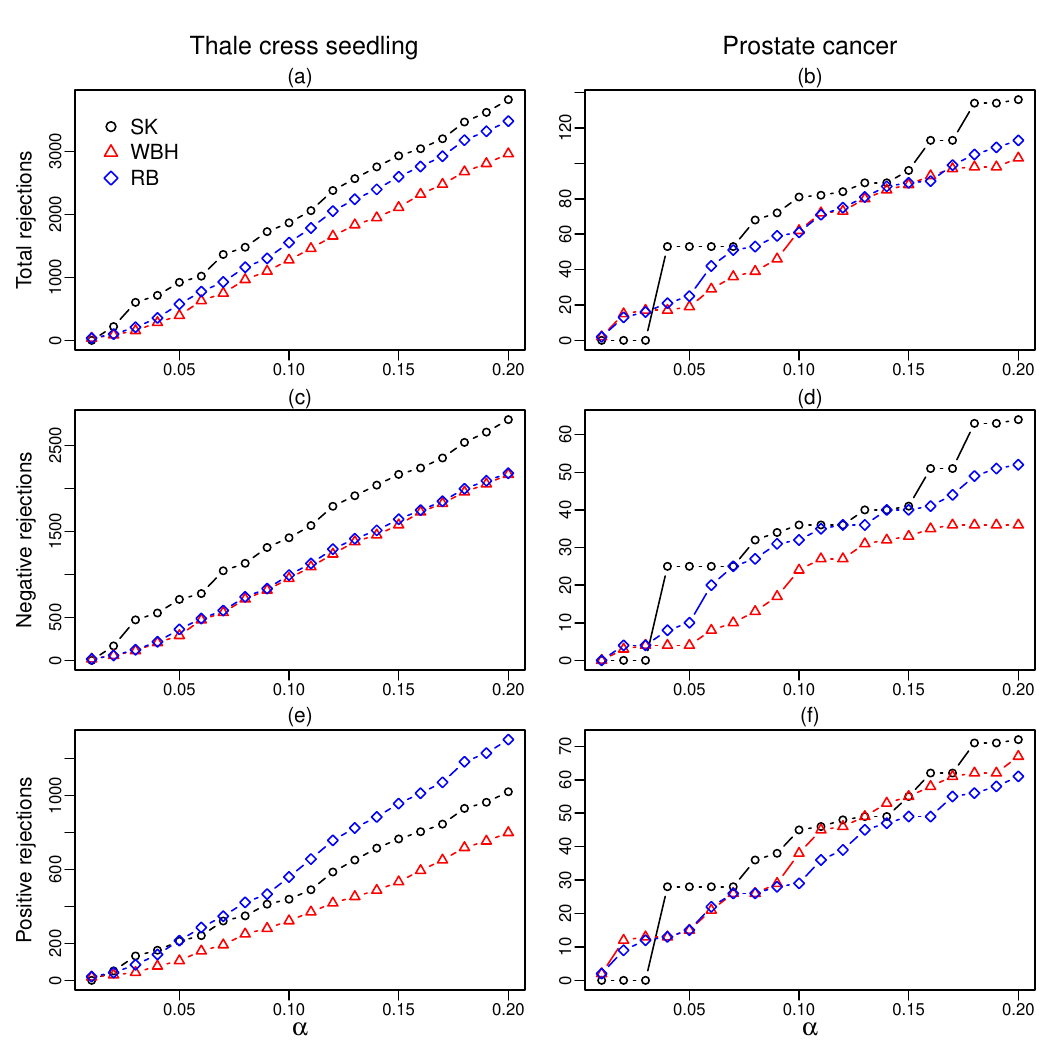}}
\caption{Plots of numbers of rejections versus the nominal FDR level $\alpha$. Panels (a), (c) and (e) show numbers of rejections in thale cress seedling data analysis, and Panels (b), (d) and (f) show numbers of rejections in prostate cancer data analysis. The first row shows the total numbers of rejections, the second row shows the numbers of rejections on the negative side, and the last row shows the numbers of rejections on the positive side.}
\label{fdrfig10}
\end{figure}

Panel (a) of Figure~\ref{fdrfig9} shows the histogram of test statistics as well as the rejection boundaries of all procedures considered when the target FDR level $\alpha=0.1$.
The $t$-statistic distribution is highly asymmetric.
There are 646 $t$-statistics that are less than $-5$, while there are only 378 $t$-statistics that are greater than $5$.
We use the Kolmogorov-Smirnov test to test for any distributional difference between the positive and negative $t$-statistics, and the $p$-value is almost zero ($<2.2$e$-16$).
As a $p$-value based method, \texttt{RB} cannot utilize the directional information and report a symmetric rejection region about 0.
On the other hand, the directional-adaptive procedures \texttt{SK} and \texttt{WBH} can use asymmetric rejection boundaries to gain power.
The distribution of test statistics is left-skewed, and both of the left boundaries of rejection regions from \texttt{SK} and \texttt{WBH} are closer to 0 than their corresponding right boundaries.
The panels in the left column of Figure~\ref{fdrfig10} show the relationship between the numbers of rejections and the nominal FDR levels.
\texttt{SK} has the most numbers of rejections overall and on the negative side, while \texttt{RB} has the most rejections on the positive side.
In essence, by using asymmetric rejection boundaries, \texttt{SK} can improve the overall power by prioritizing rejections from the less ``risky'' negative side.
On the other hand, \texttt{WBH} has fewer rejections than \texttt{RB} overall and on the positive side, and its rejections on the negative side are roughly the same as that of \texttt{RB}.

\subsection{Prostate cancer}
\cite{Singh02} conducted a microarray study to identify DE genes between prostate cancer patients and healthy individuals.
The expression levels of $n=6033$ genes are measured on 52 prostate cancer patients and 50 healthy control subjects.
We perform the common two-sample $t$-test for each gene and obtain a list of $t$-statistics, $t_1, \ldots, t_n$, and the corresponding $p$-values, $p_1, \ldots, p_n$.

Panel (b) of Figure~\ref{fdrfig9} shows the histogram of test statistics as well as the rejection boundaries of all procedures considered when the nominal FDR level $\alpha=0.1$.
There is no obvious distributional asymmetry, and the $p$-value from the test of symmetry using the Kolmogorov-Smirnov test is about 0.4.
Panels in the right column of  Figure~\ref{fdrfig10} show the relationship between the numbers of rejections and the nominal FDR levels.
For reasonably large $\alpha$, more specifically, when $\alpha>0.03$, \texttt{SK} has the most overall rejections as well as the rejections on the positive side or on the negative side for nearly all $\alpha$ values.
Also in panel (b) of Figure~\ref{fdrfig9}, the rejection region from \texttt{SK} contains the other two rejection regions of \texttt{WBH} and \texttt{RB}.
If we look more closely, there is a subtle asymmetry between the positive and negative sides.
It can be seen from the histogram in panel (b) of Figure~\ref{fdrfig9}, the positive $t$-statistics has a longer tail than that of the negative ones, indicating stronger effect sizes of up-regulated genes.
As a result, we can see from Figure~\ref{fdrfig9} panel (b) that both of the directional-adaptive procedures \texttt{SK} and \texttt{WBH} have more rejections on the positive side than on the negative side.
This example shows that our method \texttt{SK} can have equal or better power than the $p$-value based procedure \texttt{RB} even when the directional information is not strong.

\section{Conclusions and future work}
\label{sec6}
For multiple hypothesis testing problems with directional information, which are common in scientific research,
we propose the first direction-adaptive procedure that controls the FDR in finite samples.
Simulation results demonstrate our power advantage over existing procedures.

Though our theory is established under the independence condition, we show through simulation that our procedure is reasonably robust in a setting that mimics the dependence pattern in gene expression data.
The theoretical and numerical studies of other dependence settings are left for future research.




\bibliographystyle{biom}
\bibliography{reff}

\newcommand{\noop}[1]{}
\begin{thebibliography}{}

\bibitem[\protect\citeauthoryear{Barber and Cand{\`e}s}{Barber and
  Cand{\`e}s}{2015}]{Barber15}
Barber, R.~F. and Cand{\`e}s, E.~J. (2015).
\newblock Controlling the false discovery rate via knockoffs.
\newblock {\em The Annals of Statistics} {\bf 43,} 2055--2085.

\bibitem[\protect\citeauthoryear{Benjamini and Hochberg}{Benjamini and
  Hochberg}{1995}]{Ben95}
Benjamini, Y. and Hochberg, Y. (1995).
\newblock Controlling the false discovery rate: a practical and powerful
  approach to multiple testing.
\newblock {\em Journal of the Royal Statistical Society: Series B} {\bf 57,}
  289--300.

\bibitem[\protect\citeauthoryear{Benjamini and Hochberg}{Benjamini and
  Hochberg}{2000}]{Ben00}
Benjamini, Y. and Hochberg, Y. (2000).
\newblock On the adaptive control of the false discovery rate in multiple
  testing with independent statistics.
\newblock {\em Journal of Educational and Behavioral Statistics} {\bf 25,}
  60--83.

\bibitem[\protect\citeauthoryear{Efron, Tibshirani, Storey, and Tusher}{Efron
  et~al.}{2001}]{Efron01}
Efron, B., Tibshirani, R., Storey, J., and Tusher, V. (2001).
\newblock {Empirical {Bayes} analysis of a microarray experiment}.
\newblock {\em Journal of the American Statistical Association} {\bf 96,}
  1151--1160.

\bibitem[\protect\citeauthoryear{Jang, Park, Lee, Thu, Kim, Suh, Kang, and
  Kim}{Jang et~al.}{2014}]{Jang14}
Jang, Y.~H., Park, H.-Y., Lee, K.~C., Thu, M.~P., Kim, S.-K., Suh, M.~C., Kang,
  H., and Kim, J.-K. (2014).
\newblock A homolog of splicing factor {SF1} is essential for development and
  is involved in the alternative splicing of {pre-mRNA} in {Arabidopsis}
  thaliana.
\newblock {\em The Plant Journal} {\bf 78,} 591--603.

\bibitem[\protect\citeauthoryear{Jin and Cai}{Jin and Cai}{2007}]{Jin07}
Jin, J. and Cai, T.~T. (2007).
\newblock Estimating the null and the proportion of nonnull effects in
  large-scale multiple comparisons.
\newblock {\em Journal of the American Statistical Association} {\bf 102,}
  495--506.

\bibitem[\protect\citeauthoryear{Lei and Fithian}{Lei and
  Fithian}{2018}]{Lei18}
Lei, L. and Fithian, W. (2018).
\newblock {AdaPT}: an interactive procedure for multiple testing with side
  information.
\newblock {\em Journal of the Royal Statistical Society: Series B} {\bf 80,}
  649--679.

\bibitem[\protect\citeauthoryear{Liang}{Liang}{2016}]{Liang16}
Liang, K. (2016).
\newblock False discovery rate estimation for large-scale homogeneous discrete
  p-values.
\newblock {\em Biometrics} {\bf 72,} 639--648.

\bibitem[\protect\citeauthoryear{Liang and Nettleton}{Liang and
  Nettleton}{2012}]{Liang12.FDR.Est}
Liang, K. and Nettleton, D. (2012).
\newblock Adaptive and dynamic adaptive procedures for false discovery rate
  control and estimation.
\newblock {\em Journal of the Royal Statistical Society, Series B} {\bf 74,}
  163--182.

\bibitem[\protect\citeauthoryear{MacDonald and Liang}{MacDonald and
  Liang}{2017}]{MacDonald17}
MacDonald, P. and Liang, K. (2017).
\newblock Dynamic adaptive procedures that control the false discovery rate.
\newblock {\em arXiv preprint arXiv:1712.02043} .

\bibitem[\protect\citeauthoryear{M{\"u}ller, Parmigiani, Robert, and
  Rousseau}{M{\"u}ller et~al.}{2004}]{Muller04}
M{\"u}ller, P., Parmigiani, G., Robert, C., and Rousseau, J. (2004).
\newblock Optimal sample size for multiple testing: the case of gene expression
  microarrays.
\newblock {\em Journal of the American Statistical Association} {\bf 99,}
  990--1001.

\bibitem[\protect\citeauthoryear{Orr, Liu, and Nettleton}{Orr
  et~al.}{2014}]{Orr14}
Orr, M., Liu, P., and Nettleton, D. (2014).
\newblock An improved method for computing q-values when the distribution of
  effect sizes is asymmetric.
\newblock {\em Bioinformatics} {\bf 30,} 3044--3053.

\bibitem[\protect\citeauthoryear{Singh, Febbo, Ross, Jackson, Manola, Ladd,
  Tamayo, Renshaw, D'Amico, Richie, Lander, Loda, Kantoff, Golub, and
  Sellers}{Singh et~al.}{2002}]{Singh02}
Singh, D., Febbo, P.~G., Ross, K., Jackson, D.~G., Manola, J., Ladd, C.,
  Tamayo, P., Renshaw, A.~A., D'Amico, A.~V., Richie, J.~P., Lander, E.~S.,
  Loda, M., Kantoff, P.~W., Golub, T.~R., and Sellers, W.~R. (2002).
\newblock Gene expression correlates of clinical prostate cancer behavior.
\newblock {\em Cancer Cell} {\bf 1,} 203--209.

\bibitem[\protect\citeauthoryear{Storey, Taylor, and Siegmund}{Storey
  et~al.}{2004}]{Storey04}
Storey, J.~D., Taylor, J.~E., and Siegmund, D. (2004).
\newblock Strong control, conservative point estimation and simultaneous
  conservative consistency of false discovery rates: a unified approach.
\newblock {\em Journal of the Royal Statistical Society: Series B} {\bf 66,}
  187--205.

\bibitem[\protect\citeauthoryear{Sun, Craiu, Paterson, and Bull}{Sun
  et~al.}{2006}]{Sun06}
Sun, L., Craiu, R.~V., Paterson, A.~D., and Bull, S.~B. (2006).
\newblock Stratified false discovery control for large-scale hypothesis testing
  with application to genome-wide association studies.
\newblock {\em Genetic Epidemiology} {\bf 30,} 519--530.

\bibitem[\protect\citeauthoryear{Sun and Cai}{Sun and Cai}{2007}]{Sun07}
Sun, W. and Cai, T.~T. (2007).
\newblock Oracle and adaptive compound decision rules for false discovery rate
  control.
\newblock {\em Journal of the American Statistical Association} {\bf 102,}
  901--912.

\bibitem[\protect\citeauthoryear{Zhao and Fung}{Zhao and Fung}{2016}]{Zhao16}
Zhao, H. and Fung, W.~K. (2016).
\newblock A powerful {FDR} control procedure for multiple hypotheses.
\newblock {\em Computational Statistics \& Data Analysis} {\bf 98,} 60--70.

\end{thebibliography}
\end{document}